\documentstyle[titlepage,fleqn,12pt]{article}
\begin{document}
\begin{titlepage}
\title{Symmetries in a Polynomial Formulation of the
Non-Linear $\sigma$-Model}
\author{C. D. Fosco\\ \\
University of Oxford\\
Department of Physics, Theoretical Physics\\
1 Keble Road, Oxford OX1 3NP, UK \\ \\
and\\ \\
R. C. Trinchero\\ \\
Centro At\'omico Bariloche and Instituto Balseiro\\
8400 - S. C. de Bariloche\\
Argentina}
\vspace{1cm}

\baselineskip=21.5pt
\begin{abstract}
We study the realisation of global symmetries in a polynomial
(or `linearized')
formulation of the non-linear $\sigma$-model.
We show that there are global symmetries whose corresponding
Noether currents are the
topological currents in the usual formulation. The usual (Noether)
conserved
currents associated with the internal symmetry  group are reproduced,
but part of them become non-local in terms of the dynamical variables
of the polynomial formulation.

\vskip 2cm

\vskip 0.7cm

Submitted to: Physics Letters B

\end{abstract}
\maketitle
\end{titlepage}
\baselineskip=21.5pt
\parskip=3pt
Among the many interesting properties of the non-linear
$\sigma$-model~\cite{gell},
not the least important is the fact that the dynamical variables belong
to a manifold with non-trivial geometry~\cite{zinn}, thus providing a
non-linear
realisation~\cite{gasi} of the corresponding symmetry group.
As a consequence, either the Lagrangian becomes non-polynomial in terms
of unconstrained variables, or it becomes polynomial but in terms of
variables which satisfy a non-linear constraint. Sometimes it is
desirable to have a polynomial or `linearized' representation of the
model,
i.e. an equivalent description where the symmetry is
linearly realised, because then the constraints imposed by the Ward identities
on the counterterms are more easy to deal with~\cite{zinn}.
It is then natural to wonder what becomes, in the polynomial formulation,
of the topological currents in the non-linear formulation, whose existence
depends essentially on the non-linear constraint. We shall show that in the
polynomial formulation there are global symmetries whose Noether currents
correspond to the topological currents of the non-linear formulation.
The situation is thus somewhat analogous to that in the Sine-Gordon
Thirring correspondence, where the topological current in the former
is mapped to a Noether current in the latter.

We begin by briefly outlining how to obtain a polynomial representation.
There is a simple way to do that for the case of the $O(N)$
models, where the field is a $N$-component vector, constrained to have
constant modulus. In this case, a polynomial representation is
constructed by introducing a Lagrange multiplier for that constraint.
Note, however, that this is not trivially generalizable to the general
case, where the manifold is defined by a more complex constraint,
like the $SU(N)$ groups.
Here we will instead use the approach of refs.\cite{cart} and \cite{thie},
where a polynomial
representation for the $SU(N)$ model
in 3+1 dimensions was obtained.
The usual presentation~\cite{slav} of this model is in terms of an $SU(N)$
field
$U(x)$, with the Lagrangian
\begin{equation}
{\cal L} \,=\, \frac{1}{2} g^2 tr(\partial_{\mu}U^{\dag} \partial^{\mu}U)\;\;,
\label{001}
\end{equation}
where $g$ is a coupling constant with dimensions of mass (the constant
$f_{\pi}$ in its application to Chiral Perturbation Theory). The
polynomial description~\cite{thie,cart} of this model was constructed in
terms of a
non-Abelian ($SU(N)$) vector field $L_{\mu}$ and a non-Abelian
antisymmetric tensor field $\theta_{\mu \nu}$, with the Lagrangian
\begin{equation}
{\cal L} \,=\, \frac{1}{2} g^2 tr (L_{\mu} L^{\mu}) +
g \, tr (\theta_{\mu \nu} F^{\mu \nu}(L) )
\label{002}
\end{equation}
where $F_{\mu \nu} = \partial_{\mu}L_{\nu} - \partial_{\nu} L_{\mu}
+[L_{\mu},L_{\nu}]$.
The Lagrange multiplier $\theta_{\mu \nu}$ imposes the constraint
$F_{\mu \nu}(L) =0$, which is equivalent~\cite{itzy} to
$L_{\mu} = U\partial_{\mu}
U^{\dag}$, where $U$ is an element of $SU(N)$. When this is substituted back
in (\ref{002}), (\ref{001}) is
obtained \footnote{For a complete derivation of the
equivalence between the theories defined by (\ref{001}) and (\ref{002})
within the path integral
framework, see ref~\cite{cart}.}. Note that
one can always find locally a $U$ such that $L_{\mu}= U \partial_{\mu}
U^{\dag}$ if $F_{\mu \nu} =0$, but of course in general more than one
coordinate chart will be necesary to specify $U$ completely.

There
is a technical problem when the
spacetime dimension is higher than 2: there is a gauge symmetry
under transformations of the antisymmetric field~\cite{anti}. We will not
reproduce here the gauge fixing procedure (discussed in
ref.~\cite{cart}), because we will discuss only global symmetries of
the antisymmetric field,
and it is clear that it is not necesary to fix the
gauge in order to study them or their corresponding charges.

Before proceeding, we note that the essential difference
with the usual linearization approach (the one used in the $O(N)$ models)
is that the field $U$ does not
appear explicitly in the Lagrangian (this is not a problem regarding
matrix elements for the scattering of `pions', because they can be obtained
by applying a reduction process to the Green's functions with $L_{\mu}$
legs~\cite{slav}).
However, it is still possible to relate off shell Green's functions
of the field $U$ to the corresponding ones of the field
$L_{\mu}$. One should only realise that
\begin{equation}
L_{\mu} \,=\, U\partial_{\mu} U^{\dag} \Rightarrow D_{\mu}U =0, \;\;
D_{\mu} \,\equiv \, \partial_{\mu} + L_{\mu} \;,
\label{003}
\end{equation}
and then $U$ can be obtained at the point $x$ by parallel transporting
its value at spatial infinity, which we fix to be equal to the unit
matrix\footnote{We identify (as usual) all the points at spatial infinity.}:
\begin{equation}
U(x)\,=\, {\cal P} \exp [-\int_{C_x} dy^{\mu} L_{\mu} (y) ],
\label{004}
\end{equation}
where ${\cal P}$ is the path-ordering operator~\cite{itzy}, and the
line-integral
in the exponent is over a curve $C_x$, which is any regular path starting
at spatial infinity, and ending at $x$. Clearly, the condition $F =0$
guarantees that $U$ is in fact invariant under deformations of $C_x$
which leave its endpoints unchanged. Note that spatial infinity
is a unique point, and that its time argument is irrelevant because
of the time-independence of the boundary conditions.
We can also construct products of two or more fields
in a similar way, for example
\begin{equation}
U(x_2) \, U^{-1} (x_1) \;=\;{\cal P} \exp [ - \int_{C_{x \to y}}
dy_{\mu}
L^{\mu} (y) ] \;,
\label{0041}
\end{equation}
where $C_{x \to y}$ is a continuous path from $x_1$ to $x_2$. This
shows how $U$ field correlation functions can in principle be
calculated using only the $L_{\mu}$ field Lagrangian (\ref{002});
one has to evaluate, for example, the Wilson line (\ref{0041}) in
the theory defined by (\ref{002}).
We also see here a qualitative difference between this formulation
and the usual one. It will show up whenever the manifold where
the fields are defined admits any loop not deformable to a point.
In this case, we can write the lhs of (\ref{0041}) using two different
paths $C_{x_1 \to x_2}$ and ${C'}_{x_1 \to x_2}$ in the rhs.
As  the result should be the same for different paths, we obtain
the condition:
\begin{equation}
{\cal P} \{ \exp [ - \oint_C d y^{\mu} L_{\mu} (y) ] \}\;=\; 1 \;
\end{equation}
where $C \,=\, C_{x_1 \to x_2}\; {C'}^{-1}_{x_1 \to x_2}$.
This imposes a quantization condition on the flux, which can be
non-zero even when $F=0$ locally. The non-trivial topology can appear
not only because of some {\em a priori} non simply-connected
spacetime manifold. It is also present when we have vortices, as
in the $U(1)$ model in 2 Euclidean dimensions. There the manifold
has holes at the vortices' centers, because they should be excluded
in order to keep the action finite, and also because the spin is
ill-defined there. In this case, (6) implies that the flux
of each vortex is an integer multiple of $2\pi$.

Let us show now how a global symmetry in the polynomial theory
can correspond  to a topological
current in the non-linear theory.
The simplest non-trivial example is the $U(1)$ model in 1+1 dimensions;
in this case the Lagrangian (\ref{002}) can be written as
\begin{equation}
{\cal L} \,=\, \frac{1}{2} \, g^2 \, L_{\mu} L^{\mu} + \frac{1}{2}
g \theta \epsilon^{\mu \nu} F_{\mu \nu} (L) \;,
\label{005}
\end{equation}
where we have taken advantage of the fact that in 1+1 dimensions
one can write any antisymmetric $\theta_{\mu \nu}$ as
$\theta_{\mu \nu} = \frac{1}{2} \, \theta \, \epsilon_{\mu \nu}$,
where $\theta$ is a pseudoscalar field. Note then that under the
global transformation:
\begin{equation}
\theta (x) \to \theta (x) + \alpha \;,
\label{006}
\end{equation}
where $\alpha$ is a constant, the corresponding variation in
${\cal L}$ is a total derivative:
\begin{equation}
\delta {\cal L} \,=\, \partial_{\mu}  \Lambda^{\mu} \;,\;
\Lambda^{\mu} = \alpha \, g \, \epsilon^{\mu \nu} L_{\nu} \;,
\label{007}
\end{equation}
and it is straightforward to see that the corresponding Noether
current and charge are, respectively:
\begin{equation}
J^{\mu} = \frac{g}{2\pi} \epsilon^{\mu \nu} L_{\nu} \;,\;
Q = \frac{1}{2\pi} \int_{-\infty}^{+\infty} dx^1 L_1 (x^0,x^1) \;.
\label{008}
\end{equation}
The fact that $Q$ is a topological invariant is already seen from
(\ref{008}) and $F_{\mu \nu} = 0$, because it is the integral
over a one dimensional manifold of a closed 1-form (De Rham's
theorems~\cite{flan}).
If we wish to compare this charge with the usual expression in terms
of an (in general multivalued) angle, we realise that $F_{\mu \nu}=0$ is
equivalent
to $L_{\mu} = \frac{1}{g}
\partial_{\mu}
\phi$ (where $\phi$ is the angle which determines the direction of
the continuum spin, and the $\frac{1}{g}$ factor is in order to have standard
dimensions for $L_{\mu}$ and $\phi$), and so $Q$ becomes
\begin{equation}
Q \,=\, \frac{1}{2 \pi} \int_{-\infty}^{+\infty} dx^1 \partial_1
\phi (x^0,x^1) = N \;,
\label{009}
\end{equation}
where N clearly measures the winding number of the mappings from
the (assumed periodic) space $S^1$ to the spin space $S^1$. And
this result coincides with the usual topological charge of the model,
but note that in the standard treatment the topological current is
not a Noether current, but must instead be constructed using
topological arguments. Its conservation is deduced from the fact that
the time evolution is a continuous deformation of the configuration,
and the winding number is invariant under such a transformation.

The next example we consider is the group $SU(2)$ in 3+1
dimensions. The relevant global symmetry transformation is
also a transformation of the Lagrange multiplier only:
\begin{equation}
\delta \theta_{\mu \nu} \,=\, \alpha \, {\tilde F}_{\mu \nu} (L)
\label{010}
\end{equation}
where ${\tilde F}_{\mu \nu} (L) = \frac{1}{2}  \epsilon_{\mu \nu
\rho \sigma} F^{\rho \sigma} (L)$.
Note that again the variation of ${\cal L}$ is a total derivative,
\begin{equation}
\delta {\cal L} \,=\, \alpha \, g \,{\tilde F}^{\mu \nu} F_{\mu \nu}
\,=\, \partial_{\mu} \Lambda^{\mu} \;.
\label{0105}
\end{equation}
The Noether construction implies that $\Lambda^{\mu}$
becomes proportional to the conserved current
$J^{\mu}$, thus
\begin{equation}
J^{\mu} \,=\, \epsilon^{\mu \nu \rho \sigma} ( L_{\nu}\partial_{\rho}
L_{\sigma} + \frac{2}{3} L_{\nu} L_{\rho} L_{\sigma})\;.
\label{011}
\end{equation}
When this current is evaluated using the constraint equation
$F^{\mu \nu} =0$, we can write it in the equivalent
forms:
\begin{eqnarray}
J^{\mu} \,&=&\, \epsilon^{\mu \nu \rho \sigma} L_{\nu} L_{\rho} L_{\sigma}
\nonumber\\
&=&\, \epsilon^{\mu \nu \rho \sigma} U\partial_{\nu} U^{-1}
U\partial_{\rho} U^{-1} U \partial_{\sigma} U^{-1}
\label{012}
\end{eqnarray}
which is the well known topological current of the model. The second
expression in (\ref{012}) was introduced to compare with the usual
non-polynomial treatment in terms of $U$, but of course one already knows
from the first equation in (\ref{012}) that $Q$ is a topological invariant,
because
it is the integral over $S^3$ of a closed
3-form.  In the
Skyrme model{\footnote{Note that the only modification to write the
Skyrme model within the polynomial formulation is to add the
stabilization
term (a function of $L_{\mu}$ only) to ${\cal L}$,
which is clearly not affected by the
transformation (\ref{010}), and the full Lagrangian becomes invariant.}}
it is (proportional to) the baryonic current, and its charge is
proportional to the winding number of the mappings from $S^3$ (space)
to $S^3$ (internal).

We now consider the realisation of the remaining symmetries, associated
with the invariance under $SU(N)$ left and right transformations.
In the usual formulation, these correspond to left and right
multiplication of $U$ by a constant $SU(N)$ matrix~\cite{slav}.
The corresponding Noether
currents are
\begin{equation}
L_{\mu} = U\partial_{\mu} U^{-1} \;\; \;\;\;\; R_{\mu} = U^{-1} \partial_{\mu}
U\;,
\label{013}
\end{equation}
and they are conserved as a consequence of the equations of motion.
In the polynomial version, the (infinitesimal) left transformation is realised
by the
following transformation of the fields:
\begin{eqnarray}
\delta L_{\mu} (x) &=&[ u , L_{\mu} (x)] \;, \nonumber \\
\delta \theta_{\mu \nu} (x) &=& [u , \theta_{\mu \nu} (x) ]
\label{014}
\end{eqnarray}
where $u$ is a constant matrix of the Lie algebra of the group.
Naturally enough, the conserved current due to this symmetry is
$L_{\mu}$, but note that {\em there is no
transformation of the fields due to a right transformation}. One might
have expected this since $R_{\mu}$ is not completely independent of
$L_{\mu}$, but they enjoy the relationships:
\begin{equation}
R_{\mu} (x) = -  U(x) \, L_{\mu} (x) \, U^{-1} (x) \; \Rightarrow
\partial_{\mu} R^{\mu} = - \, U^{-1} \, \partial_{\mu} L^{\mu} \, U\;,
\label{015}
\end{equation}
and so their conservation is not independent. Anyway, it is desirable to
have an expression for $R^{\mu}$ in terms of the fields of the
polynomial formulation alone, because we would like them to provide a
{\em complete} description of the configurations. This can be done since we
know
how to write $U$ (Equation (\ref{004})) in terms of $L_{\mu}$.
Using (\ref{004}) in (\ref{015}), the desired
expression of $R$ in terms of $L$ is obtained:
\begin{equation}
R_{\mu} (x) = - {\cal P} \{ L_{\mu} (x) \exp [ - \int_{D_x}
d y^{\mu} L_{\mu} (y) ] \} \;,
\label{018}
\end{equation}
where $D_x$ is now a regular curve which starts at $\infty$, passes through
$x$, and comes back to $\infty$. Note that in this way we have obtained
a closed expression for the right current in terms of the left one
{\em plus the curve $D_x$ and the value of U at infinity}.
An interesting picture for
the manifestation of the chiral symmetry breaking then emerges: although
$L_{\mu}$
and $\theta$ are invariant under the right transformations, $U$ is not,
because it depends on the `vacuum' value of $U$ at $\infty$ (which we
have taken to be 1), and this will change under a right transformation.
Clearly $U$ will enjoy then the same transformation properties
as in the usual formulation.
Thus, although the original fields in the polynomial formulation do not
show the existence of the chiral symmetry breaking, we could construct
a {\em non-local}  function of them plus the vacuum value of $U$ which
does that. Then the complete state of the system is characterized by
$L_{\mu}$ and $U(\infty)$ in our picture, and the vacuum is just given
by the simultaneous conditions: $L_{\mu} =0$ and $U(\infty) = U_0$,
where $U_0$ is a constant $SU(N)$ matrix. Then it cannot be invariant
under $SU(N)_L \times SU(N)_R$ and the symmetry is broken down to
the diagonal subgroup, as usual.

Particularizing to the case of 1+1 dimensions, and a non-Abelian group,
the conserved charges
associated to $L_{\mu}$ and $R_{\mu}$ are written as:
\begin{eqnarray}
Q_L \, &=& \, \int_{-\infty}^{+ \infty} dx L_0 (x) \;,
\nonumber\\
Q_R &=&- \sum_{n=0}^{\infty} \frac{{(-1)}^n}{n!}
\int_{-\infty}^{+\infty} dx
\int_{-\infty}^{+\infty} dy_1 \cdots
\int_{-\infty}^{+\infty} dy_n \nonumber \\
& & {\cal P} [L_0 (x) L_1 (y_1) \cdots L_1 (y_n) ] \;,
\label{nonl}
\end{eqnarray}
where we have only shown the space arguments, the time arguments being the
same for
all the fields. Obviously, we have taken here a path $D_x$ which starts
at $-\infty$, passes through $x$, and then goes to $-\infty$ (which is
identified with $+\infty$). ${\cal P}$ orders the fields
according to their spatial arguments only, since we have chosen a
constant-time path.
(\ref{nonl}) shows that the polynomial description
contains non-local charges. In general, there will
be an infinite number of different paths, all leading to the same charge,
because of the path-independence of $U$
{\footnote{If the path chosen for the construction of
$Q_R$ were not a constant-time one,
of
course the charge should still be the same (at least on-shell),
but
an independent proof of its conservation (i.e. not using the
path-independence property of $U$)
would be non-trivial, since $Q_R$ would have explicit
time-dependence.}}.

We end this paper by proposing two different extensions of the polynomial
formulation. The first one is to the case of the so-called $O(2N+1)$-models,
defined by the following Lagrangian  and constraint:
\begin{equation}
{\cal L} \,=\, \frac{1}{2} \partial_{\mu} \phi^t (x) \partial^{\mu} \phi
(x)
\;\;, \phi^t (x) \phi (x) = 1\;,
\label{020}
\end{equation}
where $\phi$ is a $2N+1$-component vector and $\phi^t$ its transpose.
The configuration space thus defined is the sphere $S^{2N}$, {\em which
is not a Lie group}, and so the polynomial formulation is not easily
implemented. However, remembering that this configuration space is
isomorphous to the manifold $SO(2N+1) / SO(2N)$, we can write $\phi (x)$
as:
\begin{equation}
\phi (x) \,=\, O(x)  \, v \;,
\label{021}
\end{equation}
where  $O(x)$ is a $SO(2N+1)$ field, and $v$ is a constant,
$2N+1$ component, normalized vector. It is then easy to rewrite
the original Lagrangian (\ref{020}) as
\begin{equation}
{\cal L} \,=\, \frac{1}{2} tr [ \partial_{\mu} O^t (x) \partial^{\mu}
O(x) \rho ] \;,
\label{022}
\end{equation}
where $\rho$ is the (rank-one) projector:
\begin{equation}
\rho  \,=\, v \,  v^t \;\;\; (\Rightarrow {\rho}^2 =1)\;.
\label{023}
\end{equation}
Then the  construction applied to the case of $SU(N)$ groups can be
used here, with the only difference of having a projector in the
`mass term' for the gauge field.

The final generalization goes as follows. Let us consider the Lagrangian
of Equation (\ref{002}) for the case of an Abelian group. It
describes the dynamics of a 1-form field $L = L_{\mu} dx^{\mu}$ which is
a pure gauge: $F \,=\, dL = 0$; where $d$ is the exterior
derivative operator (we use the
notation of ref.~\cite{flan}).
Then (\ref{002}) can be written as:
\begin{equation}
{\cal L} \,=\, \frac{1}{2} g^2 (L,L) + g (\theta, dL) \;,
\label{024}
\end{equation}
where $(A,B) = \frac{1}{q!} A_{\mu_1 \cdots \mu_q} B^{\mu_1 \cdots \mu_q}$
for any pair of (equal order) forms $A$, $B$. Obviously, $\theta$ is a
2-form field.
The generalization is then simply to use the Lagrangian (\ref{024}) for
a p-form field $L$, with $p > 1$.
Then $\theta$ will be a p+1 form field, because $d$ increases the order of
the form by one.
The symmetry of the antisymmetric field appears here as transformations
\begin{equation}
\theta \to \theta + \delta \omega \;,
\label{025}
\end{equation}
where $\delta$ is the adjoint of $d$ with respect to the scalar product
$(\,,\,)$.
It is easy to see that the equations of motion for (\ref{024}) imply
\begin{equation}
dL =0 \;\; \delta L =0 \;
\end{equation}
and then
\begin{equation}
\Delta L = 0 \;\;, \Delta = \frac{1}{2} (\delta d + d \delta) \;.
\label{027}
\end{equation}
The operator $\Delta$ is the Laplacian for the p-form field $L$.
Let us apply this procedure to a 2-form field $L_{\mu \nu}$. The
Lagrangian of Eq.(\ref{024}) reads in this case:
\begin{equation}
{\cal L} \,=\, \frac{1}{2} g^2 \frac{1}{2!} L_{\mu \nu} L^{\mu \nu} \,
+ \, \frac{1}{3!} \,
g \, \theta_{\mu \nu \rho} F^{\mu \nu \rho} (L) \;,
\label{028}
\end{equation}
where
\begin{equation}
F_{\mu \nu \rho} \,=\, \partial_{\mu} \theta_{\nu \rho} +
\partial_{\nu} \theta_{\rho \mu} + \partial_{\rho} \theta_{\mu \nu}
\;.
\label{029}
\end{equation}
The Lagrange multiplier imposes the Bianchi identity on $L$, and then
this can be written as the exterior derivative of some 1-form A;
i.e. $L_{\mu \nu} \,=\, \partial_{\mu} A_{\nu} - \partial_{\nu} A_{\mu}$;
so we see that the theory reduces to the Maxwell Lagrangian for the
field $A$. If we don't introduce $A$, we have a first order formulation
of the Maxwell theory.
We conclude with the construction for this example of the analogue
of the Wilson line we considered in Eq.({\ref{0041}).
As $L$ is now a 2-form, we can integrate it over a (open) surface $S$, with
boundary $\partial S$, and exponentiate it:
\begin{equation}
\Phi (S) \,=\, \exp [ i \int_S L ] \;,
\label{030}
\end{equation}
but as $L = d A$, we can apply Stokes' theorem to that integral,
and we see that $\Phi$ depends only on $\partial S$ :
\begin{equation}
\Phi (\partial S) \,=\, \exp [i \int_{\partial S} A] \;.
\label{031}
\end{equation}
The argument leading to the condition (6) can be extended to this case
also; we have only to express $\partial S$ as the boundary of two
different surfaces, and we get the condition:
\begin{equation}
\exp ( i \oint_{\tilde S} L ) \,=\, 1 \;,
\label{032}
\end{equation}
where ${\tilde S}$ is the closed surface constructed from the two
different open surfaces whose boundary is $\partial S$, and they are
patched
at their boundaries. In the case when a single point is excluded from
the spacetime manifold, (\ref{032}) leads to the Dirac quantization
condition for the monopole:
{\footnote{Clearly the electric charge can be introduced by a redefinition of
the fields $A \to e A$.}}
\begin{equation}
\oint_{\tilde S} L \,=\, 2 \pi \, \times \, integer \;.
\label{033}
\end{equation}

\section{Acknowledgements}
C. D. F. acknowledges Prof. I. J. R. Aitchison for carefully reading
the manuscript.

\end{document}